\documentclass[8.5pt,twoside,twocolumn]{article}
\usepackage[super,sort&compress,comma]{natbib} 
\usepackage{mhchem}
\usepackage{times,mathptmx}
\usepackage{sectsty}
\usepackage{balance} 

\usepackage{graphicx} %eps figures can be used instead
\usepackage{lastpage}
\usepackage[format=plain,justification=justified,singlelinecheck=false,font=small,labelfont=bf,labelsep=space]{caption} 
\usepackage{fancyhdr}
\usepackage{caption}
\DeclareCaptionFormat{myformat}{#1#2#3\hrulefill}
\usepackage{float}
\usepackage{balance}
\usepackage{xcolor}
\usepackage{url}

\begin{document}

\fancypagestyle{plain}{
%\fancyhead[L]{\includegraphics[height=8pt]{headers/LH}}
%\fancyhead[C]{\hspace{-1cm}\includegraphics[height=20pt]{headers/CH}}
%\fancyhead[R]{\includegraphics[height=10pt]{headers/RH}\vspace{-0.2cm}}
\renewcommand{\headrulewidth}{1pt}}
\renewcommand{\thefootnote}{\fnsymbol{footnote}}
\renewcommand\footnoterule{\vspace*{1pt}% 
\hrule width 3.4in height 0.4pt \vspace*{5pt}} 
\setcounter{secnumdepth}{5}

\makeatletter 
\def\subsubsection{\@startsection{subsubsection}{3}{10pt}{-1.25ex plus -1ex minus -.1ex}{0ex plus 0ex}{\normalsize\bf}} 
\def\paragraph{\@startsection{paragraph}{4}{10pt}{-1.25ex plus -1ex minus -.1ex}{0ex plus 0ex}{\normalsize\textit}} 
\renewcommand\@biblabel[1]{#1}            
\renewcommand\@makefntext[1]% 
{\noindent\makebox[0pt][r]{\@thefnmark\,}#1}
\makeatother 
\renewcommand{\figurename}{\small{Fig.}~}
\sectionfont{\large}
\subsectionfont{\normalsize} 

%\fancyfoot{}
%%\fancyfoot[LO,RE]{\vspace{-7pt}\includegraphics[height=9pt]{headers/LF}}
%%\fancyfoot[CO]{\vspace{-7.2pt}\hspace{12.2cm}\includegraphics{headers/RF}}
%%\fancyfoot[CE]{\vspace{-7.5pt}\hspace{-13.5cm}\includegraphics{headers/RF}}
\fancyfoot[RO]{\footnotesize{\sffamily{1--\pageref{LastPage} ~\textbar  \hspace{2pt}\thepage}}}
\fancyfoot[LE]{\footnotesize{\sffamily{\thepage~\textbar\hspace{3.45cm} 1--\pageref{LastPage}}}}
\fancyhead{}
\renewcommand{\headrulewidth}{1pt} 
\renewcommand{\footrulewidth}{1pt}
\setlength{\arrayrulewidth}{1pt}
\setlength{\columnsep}{6.5mm}
\setlength\bibsep{1pt}
\setlength{\parskip}{1mm plus0mm minus0mm}

\twocolumn[
  \begin{@twocolumnfalse}
\noindent\LARGE{\textbf{Swimming of bacterium \textit{Bacillus subtilis} with multiple bundles of flagella$^\dag$}}
\vspace{0.6cm}

\noindent\large{\textbf{Javad Najafi$^{\ast}$\textit{$^{a}$}, Florian Altegoer{$^{b}$}, Gert Bange{$^{b}$}, and Christian Wagner\textit{$^{ac}$}}}\vspace{0.5cm}
%Please note that \ast indicates the corresponding author(s) but no footnote text is required. 

\vspace{0.6cm}
%Please do not change this text.

\noindent \normalsize{We characterize the bundle properties for three different strains of \textit{B. subtilis} bacteria with various numbers of flagella. Our study reveals that, surprisingly, the number of bundles is independent of the number of flagella, and the formation of three bundles is always the most frequent case. We assume that this relates to the fact that different mutants have the same body length.
There is no significant difference between the bundle width and length for distinct strains, but the projected angle between the bundles increases with the flagellar number. Furthermore, we find that the swimming speed is anti-correlated with the projected angle between the bundles, and the wobbling angle between the swimming direction and cell body increases with the number of flagella. Our findings highlight the impact of geometrical properties of bacteria such as body length and bundle configuration on their motility.}
\vspace{0.5cm}
 \end{@twocolumnfalse}
  ]

\section{Introduction}
%Footnotes
\footnotetext{\dag~Electronic Supplementary Information (ESI) available: One supplementary video which shows the switching between the double and single bundle. One PDF which contains a short description of the video, four graphs indicating various measured lengths for the strains, effective aspect ratio of the cell, and correlation between some important parameters. See DOI: 10.1039/b000000x/}

%Please use \dag to cite the ESI in the main text of the article.
%If you article does not have ESI please remove the the \dag symbol from the title and the above footnotetext.

\footnotetext{\textit{$^{a}$~Experimental Physics, Saarland University, 66123 Saarbr{\"u}cken, Germany; E-mail: jnajafi.phys@gmail.com}}
\footnotetext{\textit{$^{b}$~Department of Chemistry and Center for Synthetic Microbiology, Philipps University Marburg, 35043 Marburg, Germany}}
\footnotetext{\textit{$^{c}$~Physics and Materials Science Research Unit, University of Luxembourg,
Luxembourg}}

%additional addresses can be cited as above using the lower-case letters, c, d, e... If all authors are from the same address, no letter is required

%\footnotetext{\ddag~Additional footnotes to the title and authors can be included \emph{e.g.}\ `Present address:' or `These authors contributed equally to this work' as above using the symbols: \ddag, \textsection, and \P. Please place the appropriate symbol next to the author's name and include a \texttt{\textbackslash footnotetext} entry in the the correct place in the list.}

Many microorganisms use flexible appendages, known as flagella, to move in aqueous environments. A prokaryotic flagellum is proton motive and can spin either clockwise (CW) or counterclockwise (CCW).
Peritrichous bacteria such as \textit{E. coli} have many flagella, which can make bundles and push the cells forward when they all rotate CCW. As one or more flagella rotate CW, the bundle disassociates, and the cell experiences a slow phase of motility~\cite{berg2008,turner2016,najafi2018}. These states of the bacterial motility are called 'run' and 'tumble', respectively, which play a very important role in the search and chemotactic efficiency of cells~\cite{rupprecht2016}.
While \textit{E. coli} has served as a model system for many studies, the swimming of some other species with numerous flagella, such as \textit{B. subtilis}, is less intensively studied~\cite{Li2010,turner2016,najafi2018}.
The \textit{B. subtilis} cell is longer than \textit{E. coli} and generally has at least around two folds more flagella than its gram negative counterpart, \textit{E. coli}.
Originally, it was assumed that peritrichous bacteria always form a single bundle~\cite{berg1972,berg2008}, but it was later revealed that filamentous cells of \textit{E. coli} as well as \textit{B. subtilis} cells can form several bundles at the same time~\cite{Li2010,valeriani2011,najafi2018}.
These observations have been confirmed by a boundary element method simulation of swimming cells. In these studies, an experimentally measured pitch length larger than 4 $\mu$m in the wiggling trajectories of \textit{B. subtilis} cells was connected to the formation of several bundles~\cite{YHyon2012}. Multiple bundles have also been observed in the mesoscale hydrodynamic simulation of the swarmer \textit{E. coli} cell~\cite{eisenstecken2016}.
Quantitative dynamics of swimming with several bundles have still not been experimentally studied, even though the spinning bundles are most microorganisms' main appendages to propel themselves in the low Reynolds number regime. The reason for some bacteria forming multiple bundles is still an open question. More detailed aspects in this regard are the bundles' position relative to the cell body and to each other, which affect their hydrodynamic interaction, stability, and possibly run-tumble dynamics. These parameters can significantly change the cell's motility, dispersion, and chemotactic efficiency~\cite{najafi2018}.

In this article, we use three strains of \textit{B. subtilis} bacteria with various flagellar number as model systems to characterize the dynamics of multiple bundles.
We study the probability of formation of different bundles as the flagellar number varies. Moreover, we measure the principle geometrical features of the bundles and connect them to the swimming speed and wobbling angle. Our results demonstrate that the probability of forming a certain number of bundles is independent of the number of flagella, and the formation of the three bundles occurs in over fifty percent of the cases.  
 
\begin{figure*}[t!]
	\centering
	\includegraphics[width=.85\textwidth]{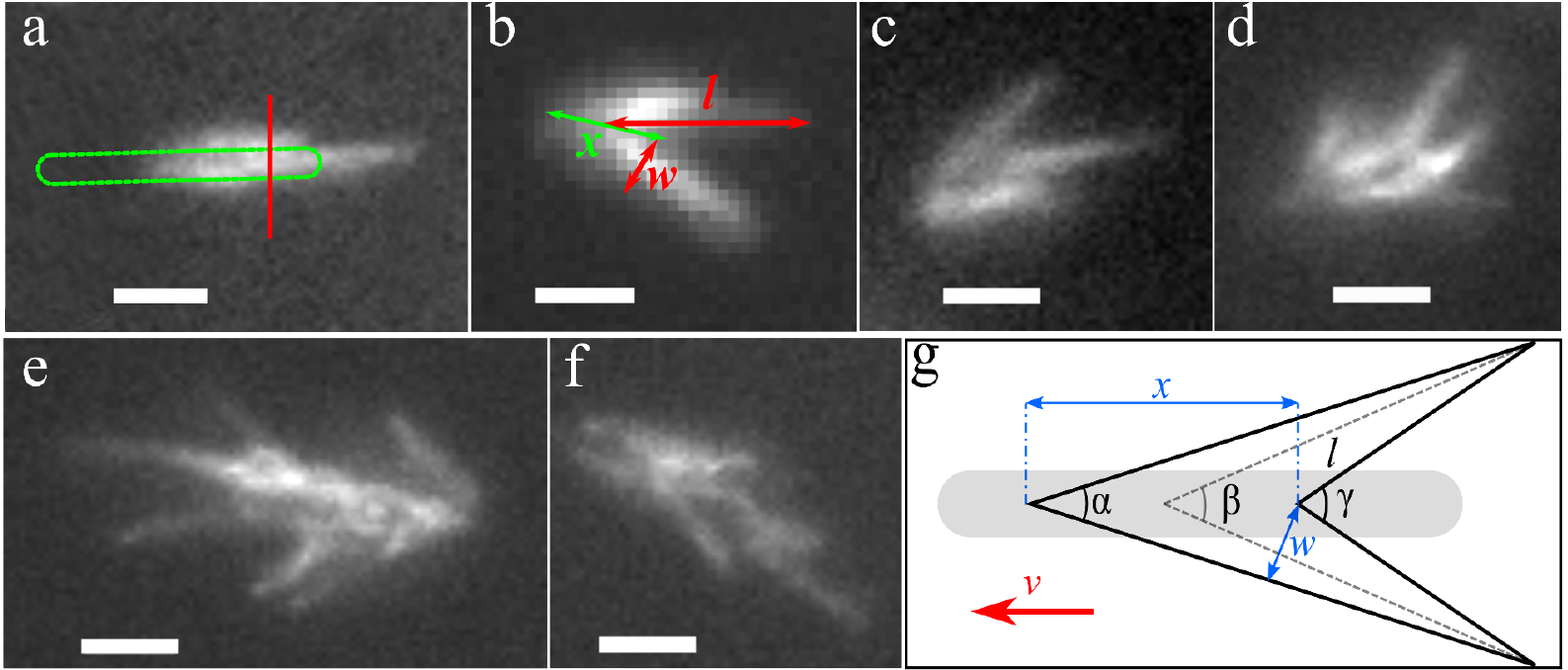}
	\caption{(a-d) Swimming \textit{B. subtilis} cells with one to four bundles. The red line in (a) represents the middle of the bundle where the bundle width is measured. The dashed green contour shows the position of the cell body. (e) Long cells form several lateral bundles, apart from the ones at the rear of cell. (f) Asymmetric bundles configuration. (g) Schematic sketch of the bundles in the observation plane and an indication of the measured quantities (see text). The key lengths are also demonstrated in (b). The illustrated rod is the approximate position of the body. Scale bar is 3.5 $\mu$m for the images.}
	\label{fig:Fluorescence}
\end{figure*}
%%_________________________________________________________________________________________________________________________________________________________________________________________________________________________
\section{Experimental Methods}

To build the strains, an allelic replacement was performed using the pMAD-system, in accordance with the literature~\cite{arnaud2004}. All the strains originated from the recently described NCIB3610 \emph{B. subtilis} strain, harboring a point mutation in the $\text\it{comI}$ gene~\cite{konkol2013}. Briefly, the $\text\it{hag}$ gene encoding flagellin was amplified, including the up- and downstream flanking regions, and cloned into the pMAD plasmid. The T209 to C mutation was obtained by quick-change mutagenesis. NCIB3610 $\text\it{comI}\!_{_\text{Q12I}}$ was transformed and positive clones selected in accordance with the previous works~\cite{arnaud2004}. The exchange of the native $\text\it{hag}$ gene with $\text\it{hag}_{_{\text{T209C}}}$ was confirmed by sequencing and light microscopy. The strain is referred to as WT NCIB3610 in terms of the flagellar number in the text. We thank DB Kearns for strain DK1693 ($\Delta\text{swrA}$, $\text\it{amyE}{::}\text{P}
\!\!_{_\text{hyspank}}\!\!{-}\text{swrA}$; $\text\it{lacA}{::}\text{P}\!_{_\text\it{hag}}\!{-}\text\it{hag}_{_{\text{T209C}}}$ mls). This 
strain was used as both $\Delta\text{swrA}$ and $\text{swrA}\!^{++}$ when induced with IPTG. The name of the strains with their corresponding number of flagella and analyzed cells in the fluorescent experiments are as follows:
\begin{table}[h!]
	\small
	\centering
	\caption{\ The three different strains with the corresponding number of flagella and analyzed cells.}
	\label{tab:strains}
	%\resizebox{\textwidth}{!}{%
		\begin{tabular}{l l l l}
			\hline
			%\hline
			Name& $\Delta$swrA & WT & swrA$^{++}$  \\
			%\hline
			\hline
			Number of flagella & 9$\pm$2  & 26$\pm$6  & 41$\pm$6  \\
			Analyzed cells & 30  & 35  & 30  \\
			\hline
	\end{tabular}
\end{table}

We streaked $\simeq$20 $\mu$l of the cryo-culture of \textit{B. subtilis} on an LB-agar plate. The plate was incubated for $\simeq$16 hours and then a few colonies of bacteria from the plate stirred in the LB to grow over night at 37 $^{\circ}$C. The cultures were diluted to optical density, measured at a wavelength of 600 nm OD$_{600}$$\simeq$0.1 the next morning and grown for two more hours to reach OD$_{600}$$\simeq$1. After dilution, we induced swrA$^{++}$ mutant by 1 mM IPTG solution to synthesize more flagella. For the fluorescent labeling of cells, we prepared the dye by dissolving 1 mg Alexa Fluor\textcircled{R}568 C5-maleimide in 200 $\mu$l DMSO (Dimethyl sulfoxide). Furthermore, we centrifuged 1 ml of the cell culture with OD$_{600}$$\simeq$1 at 8000g for one minute and gently washed it three times in PBS pH 7.4(1X) by pipetting. The pellet was resuspended in 200 $\mu$l PBS buffer along with 5 $\mu$l of the dye solution. The suspension was mixed and incubated in the dark at room temperature for 20 min, washed three times again in PBS buffer, and re-energized by half an hour outgrowth in LB to observe motile cells. To track the cell body, we grew the cultures to the optical density in the range of $\simeq$0.5 and 0.8 after dilution in the morning. We then adjusted the optical density to 0.5 and further diluted the suspension $\simeq$15 folds in the fresh LB that had already been purified by 0.4 $\mu$m syringe filter~\cite{najafi2018}. Due to the strong sticking of $\Delta$swrA strain to the surface, 0.005$\%$ PVP-40 (polyvinylpyrrolidones) was added before the experiments~\cite{berke2008}.

Fluorescence microscopy samples were prepared by adding 30 $\mu$l of the labeled cell suspension into a FluoroDish FD35-100. The drop of suspension was covered by a circular cover glass (VWR, 22 mm diameter, NO. 1.5 thickness) and then the lid of the dish was closed. A Nikon Eclipse Ti microscope, equipped with a Nikon N Plan Apo $\lambda$ 60x, N.A. 1.4 oil immersion objective, was used for the wide field fluorescence microscopy. The dye was excited by a mercury pre-centered fiber illuminator (Nikon INTENSILIGHTC-HGFIE). Videos were recorded using a Hamamatsu ORCA-Flash 4.0 camera with 25 or 30 ms exposure time and 2$\times$2 or 4$\times$4 binning. The pixel size of the camera without binning was 0.11 $\mu$m. The mirror unit was consisted of TxRed HC filter set (manufactured by Semrock). To obtain temporally long trajectories, videos were recorded in the vicinity of the glass surface, where the cells spend a long time~\cite{berke2008} and the cell density is higher~\cite{molaei2014}. The tracking chamber consisted of a cover slip (VWR, 20$\times$20 mm, NO. 1 thickness) on a microscope slide (Carl Roth GmbH, Karlsruhe, 76$\times$26 mm), separated by a thin layer of silicone grease (GE Bayer Silicones Baysilone, medium viscosity) as spacer. The chambers were quasi two dimensional, and their height was between 30 and 50 $\mu$m. The lateral sides of the chamber were sealed by the silicone grease after filling the cavity with the bacterial suspension. Afterwards, microscopy was performed by a Nikon Eclipse TE2000-s microscope and a Nikon 4x, N.A. 0.2 objective in the dark field mode. For each sample, sequences of images with 60 Hz frame rate were recorded for two minutes using a Point Grey FL3-U3-88s2cc camera. The tracking experiments were repeated with three different cultures for each strain~\cite{najafi2018}.
\begin{figure}[t!]
	\centering
	\includegraphics[width=1\linewidth]{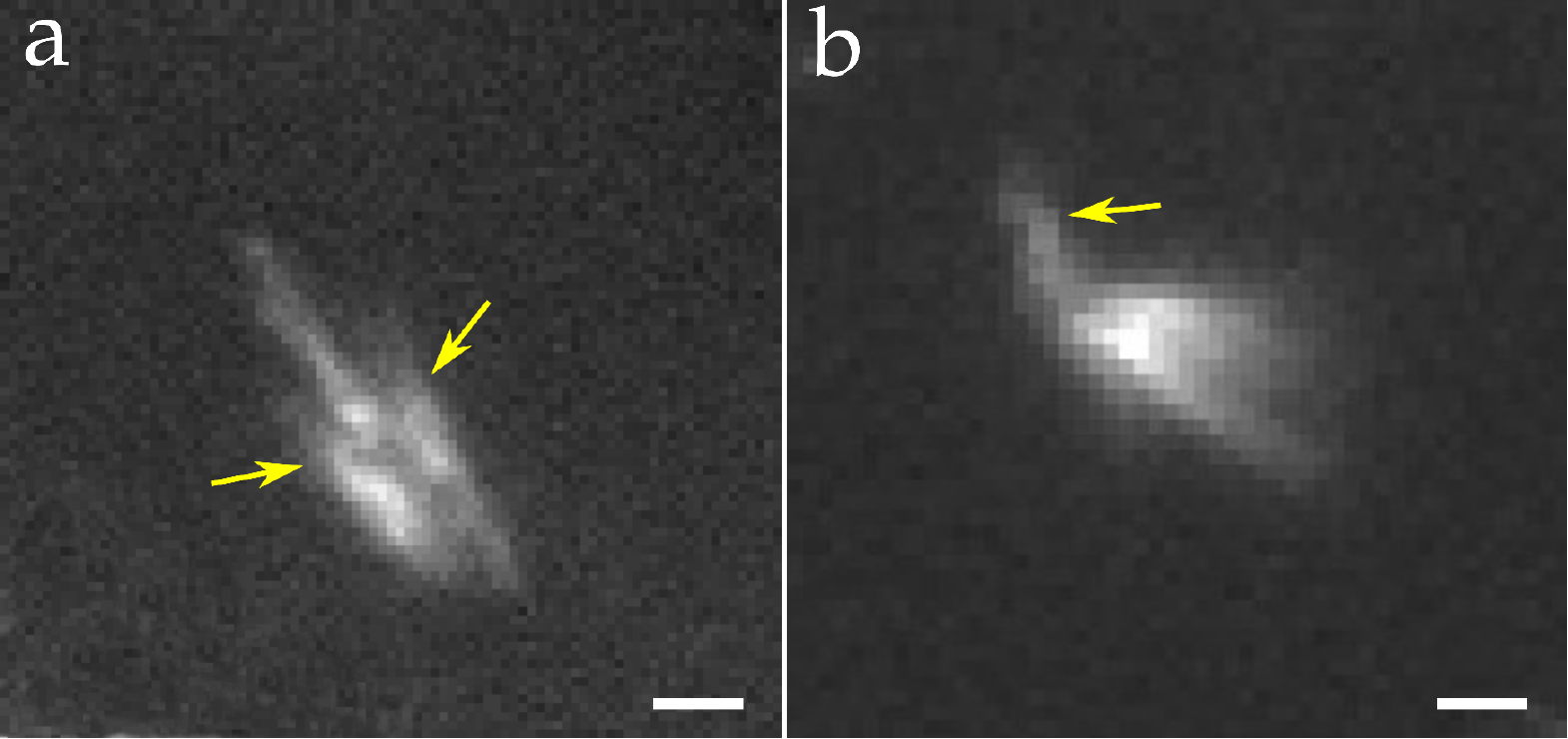}
	\caption{(a) Examples of short bundles which could not form separate effective bundles indicated by arrows. (b) Some flagella can detach from the main bundles during the swimming. Scale bars are 2 $\mu$m.}
	\label{fig:bundleextra}
\end{figure}

We analyzed the fluorescence data using Fiji package~\cite{fiji}. We considered only the regular bundles arrangements and excluded the irregular ones (Fig.~\ref{fig:Fluorescence}a-d). We assumed the arrangement of bundles as regular when the overlapping length between the bundles on the cell surface was almost maximum. In the three dimensions, they all lay down on a cone and their extensions from the base toward the cell body reaches to the same point at the apex if one neglects the widths of bundles. The number of bundles does not exceed more than four at this configuration and measuring the basic geometrical parameters are feasible for the bundles when they align on the observation plane. Irregular configurations mainly happens in the long cells where the bundles appear at different points on the cell body. Each frame of the videos was reviewed if the bundles were located on the observation plane and the intended quantities such as bundle length and width and the length of the overlapping area between the bundles and opening angles were measured (Fig.~\ref{fig:Fluorescence}g). The measurements were repeated for at least five different frames for each data point. Moreover, the labeled bundles were tracked by the TrackMate plugin of Fiji package to get the run speed after the careful exclusion of tumbling intervals. The tracking of the cell bodies was performed using a custom written script in Matlab. The trajectories were smoothed by a moving triangle and the temporally short tracks as well as the very long and slow cells were discarded~\cite{najafi2018}. We assumed the bundles always form at the rear of the cell and defined the angle between the swimming velocity and cell body centerline as the wobbling angle. To estimate the wobbling frequency, we used the oscillation of wobbling angle along the velocity direction. The amplitude of the fast Fourier transformation of wobbling angle was calculated for the trajectories which did not experience any tumbling event and the position of the global peak was considered as the wobbling frequency.

%%_____________________________________________________________________________________________________________________________________________________________________________________________________________________
\section{Results}

The first observation is that bundles can take either regular (Fig.~\ref{fig:Fluorescence}a-d) or irregular configurations (Fig.~\ref{fig:Fluorescence}e-f). Swimming cells with one long bundle at the rear have sometimes several short bundles which extend a few microns beside the cell body (Fig.~\ref{fig:bundleextra}a). The filaments within these bundles are often not long enough to form separate effective bundles and just surround the main bundle. The orientation of the main bundle remains almost constant which suggest it is aligned with the cell body and all bundles can be roughly placed on a plane. When the side flagella become a little longer, they can make two distinguishable bundles which are shorter than the main one along the cell body.
Sometimes a few flagella detach from the bundles in the run phase without perturbing them (Fig.~\ref{fig:bundleextra}b) and rejoin the bundles again. We could not determine the helicity or rotation direction of these filaments due to the insufficient resolution of images, but we suggest that these filaments rotate CW as has been reported for the single bundle of \textit{E. coli}~\cite{darnton2007}. 
Most of the bundle configurations are symmetrically aligned on a cone-like shape but sometimes their length or width are a little different. This might happen because flagella were broken by shearing or due to bending the distal end of the bundle out of the observation plane, and possibly unequal contribution of the filaments to different bundles. Cells with a long body length can form more than four bundles which are almost always asymmetric with several lateral bundles (Fig.~\ref{fig:Fluorescence}e and f). The irregular arrangements of bundles were excluded from the quantitative characterizations because we could not measure the defined parameters (Fig.~\ref{fig:Fluorescence}g). 
\begin{figure}[t!]
	\centering
	\includegraphics[width=.92\linewidth]{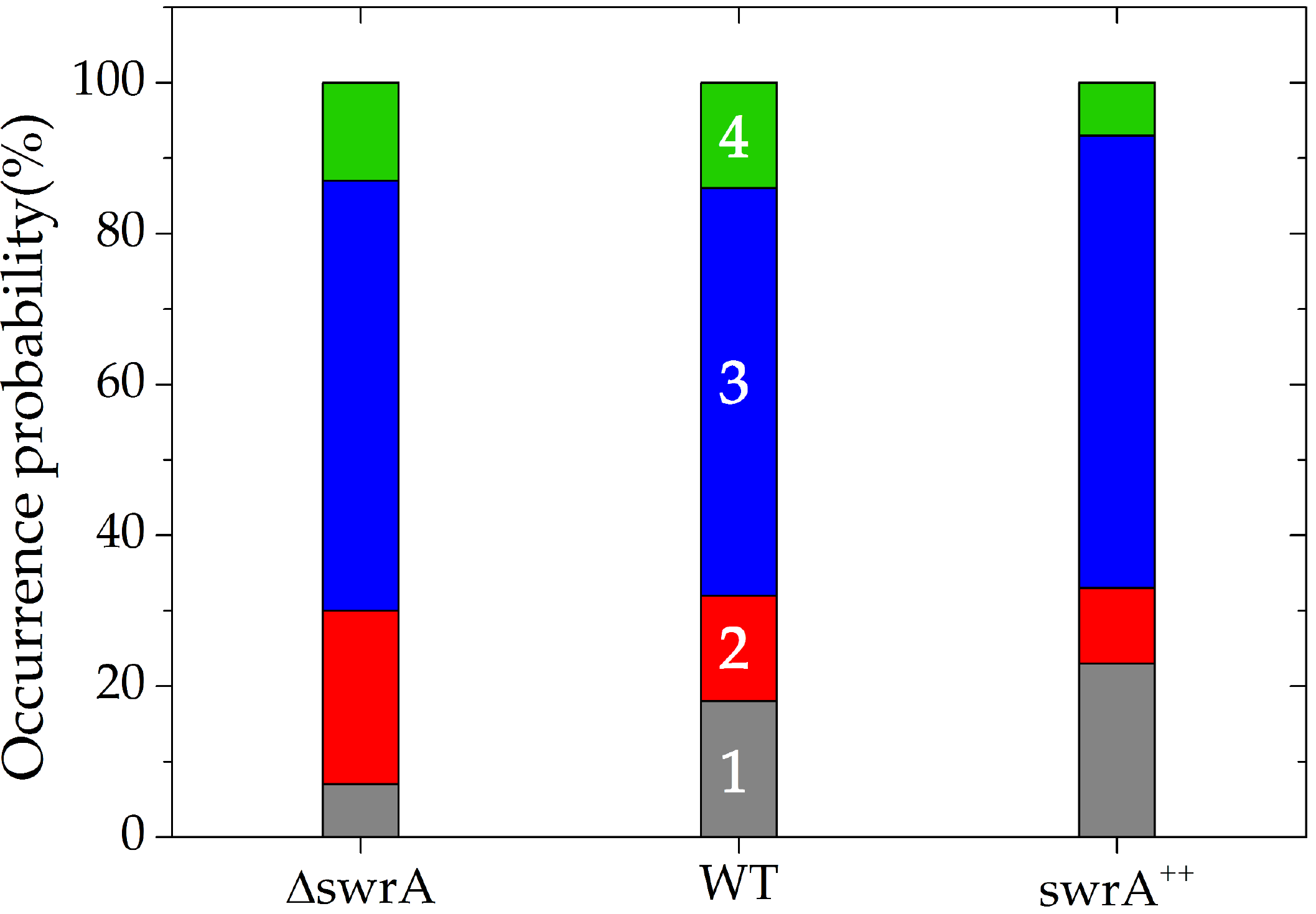}
	\caption{Probability of the number of bundles for each strain. Formation of three bundles is the most probable case regardless of the number of flagella.}
	\label{fig:Percentages}
\end{figure}

\begin{figure}[b!]
	\centering
	\includegraphics[width=1\linewidth]{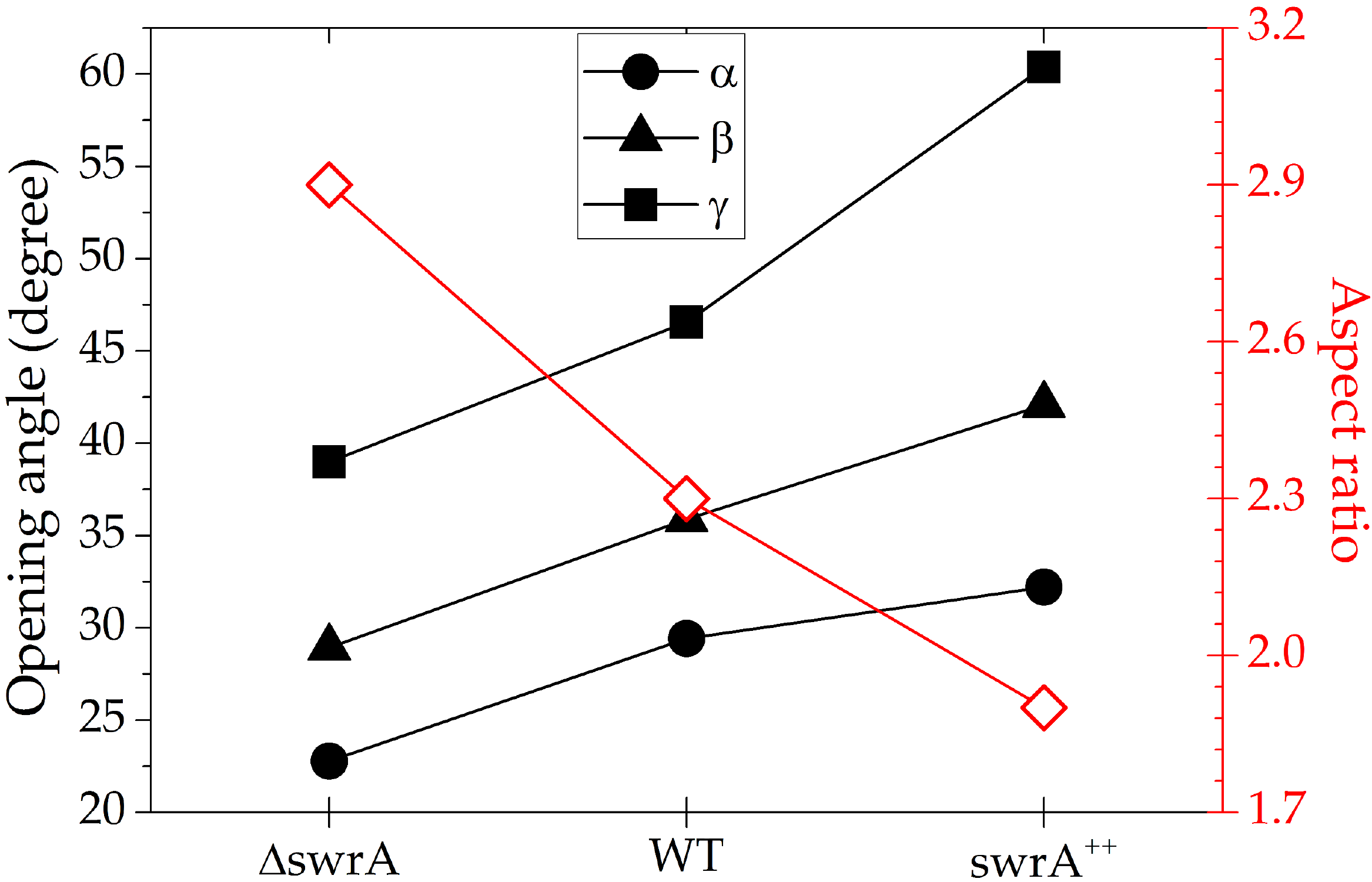}
	\caption{Projected angles between the bundles enlarge with the increase in the flagellar number. Consequently, the effective aspect ratio of the cell decreases. Exemplary aspect ratios are presented for the case in which the bundles meet each other in the middle of the cell body.}
	\label{fig:angles}
\end{figure}

\begin{figure}[t!]
	\centering
	\includegraphics[width=.92\linewidth]{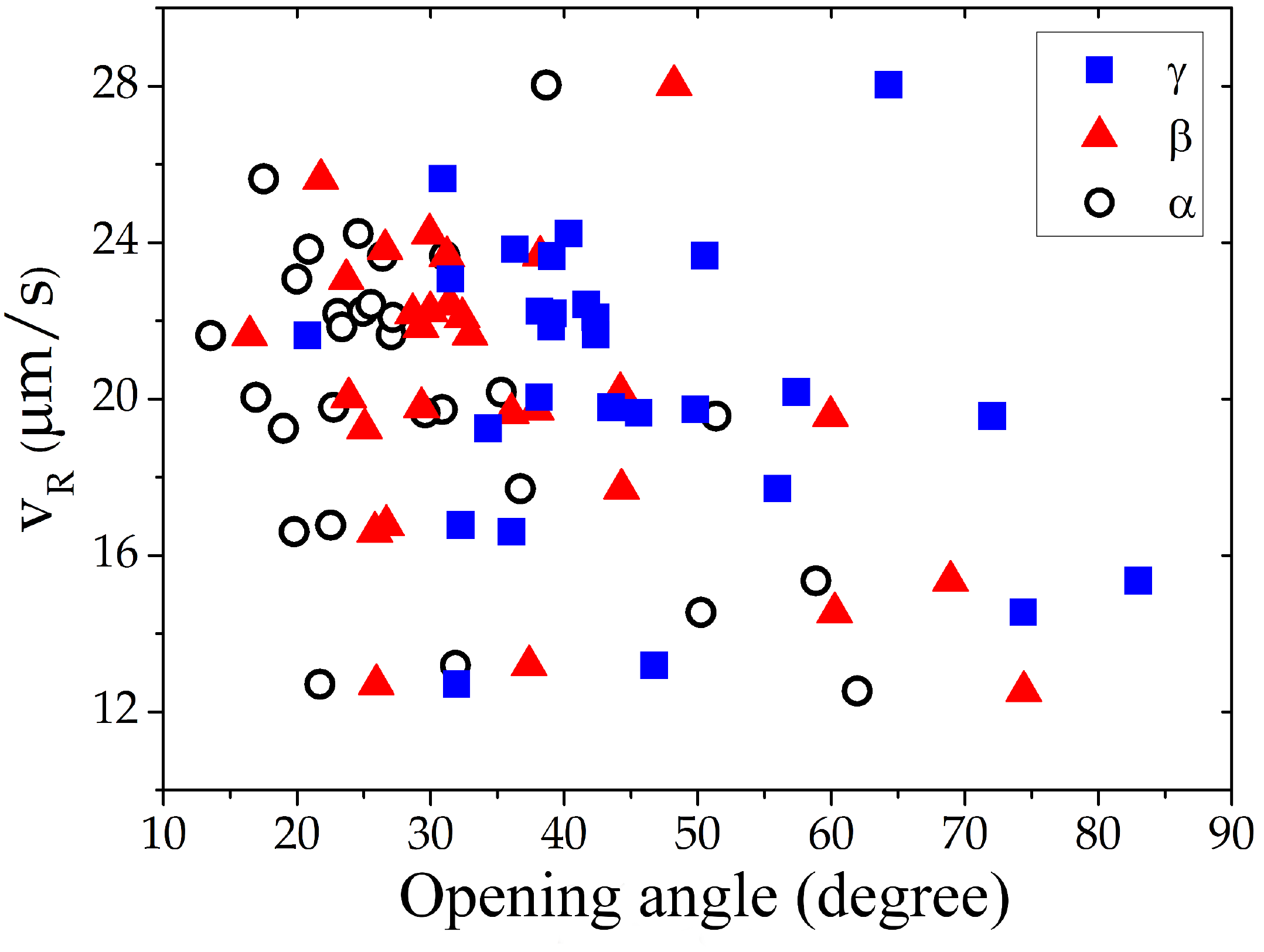}
	\caption{Swimming velocity of the wild type strain as a function of the opening angles between the bundles. The correlation coefficient for the middle angle (solid triangles) is $r{=}-0.42$. }
	\label{fig:Vr-a}
\end{figure}
\begin{figure*}[h]
	\centering
	\includegraphics[width=.9\linewidth]{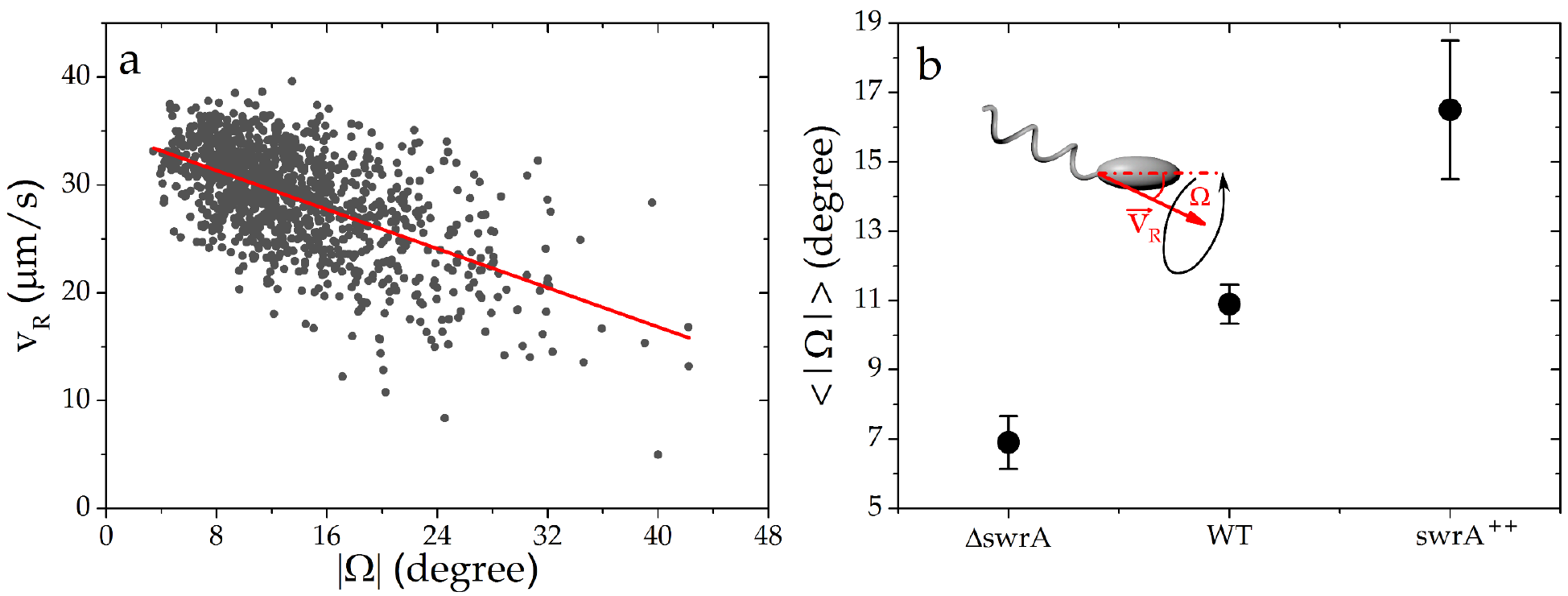}
	\caption{(a) Run speed is anti-correlated with the absolute value of the wobbling angle. The line is a linear regression fit to the data as an eye guide. The Pearson correlation coefficient is $r{=}-0.58$. (b) Wobbling angle enhances with the flagellar number. The inset image indicates how the wobbling angle is defined.}
	\label{fig:wobbling}
\end{figure*}

It was reported that long filamentous \textit{E. coli} cells have more flagella than the cells of normal size and that they form numerous bundles~\cite{Li2010}. While this effect was assigned to the number of flagella, it was not possible to determine a clear correlation between the number of flagella and bundles. By a way of contrast, it appears from our experiments that the number of bundles are determined by the cell length instead of the flagellar number (cp. Fig.~\ref{fig:Fluorescence}e). The most probable number of bundles that is formed is three, independent of the number of flagella (Fig.~\ref{fig:Percentages}). In fact, for each strain, $\sim$60$\%$ of the investigated cells have three bundles. About 20$\%$ of the studied WT and swrA$^{++}$ strains form only one bundle, which is even higher than that for the mutant with the lowest number of flagella. A speculative explanation for this could be that the effective size of the strains with a larger number of flagella is larger, and it is likely that their flagella experience more shear during pipetting, which can increase flagella breaking and result in only one bundle formation. Interestingly, the number of bundles that a certain bacterium forms is not always the same, but can vary when the bundle re-form after a run and tumble event (\textit{see Suppl. Video}). The number of bundles seemingly depends mostly on the location and number of the filaments, which contribute to each bundle. Certain filaments can go to different bundles, even if it is not clear what determines this choice. As preferably, several lateral bundles are observed in the long cells, it appears that the number of created bundles is mostly a geometrical issue, and multiple bundles are formed if not all flagella fit within a single bundle because they are too far apart from each other. Consequently, the probability of forming several bundles for all strains is similar, because their body and filaments lengths are very similar as well.

We measured the length and width of the individual bundles along with the length of the overlapping area between the bundles, which are indicated by $l$, $w$, and $x$, respectively, in Fig.~\ref{fig:Fluorescence}g. These quantities do not show any remarkable tendency and are presented in Fig. S1 in the supplementary material. The average bundle length is $\simeq$7.21 $\mu$m in agreement with the reported values in the literature~\cite{Li2010,ito2005}. The average width of a single bundle is larger than the width of the individual bundles in multiple bundle configurations. The width of the single bundles was measured in their middle, where in all probability the cell body prohibits the formation of tight bundles, even though it is expected that one single bundle should be wider.  
The average bundle width is $\simeq$1.64 $\mu$m, i.e., roughly twice the width of the cell body representing a very loose bundle. Loose bundles can incorporate some further flagella in WT strain without a noticeable variation of width; however, by placing more flagella in swrA$^{++}$, the bundle width slightly increases.
The bundle width can be modified by the flagellar rotation rate, which influences the hydrodynamic interaction, and the number of filaments that participate in the bundle formation. The geometrical restriction as well as inter-flagellar viscous drag in swrA$^{++}$ mutant are stronger because of the involvement of more flagella in the bundle, which consequently, can reduce the bundle rotation frequency and torque.
The length of the overlapping area of the bundles $x$ is, on average, about 4 $\mu$m, i.e., approximately half the length of the cell body, measured by the tracking experiments. This indicates that the overlapping area of different bundles partially covers the cell surface.  
We also measured the characteristic length of twenty stationary helical filaments of WT strain. The lengths are the pitch length $\lambda{=}2.26\pm0.17$ $\mu$m, diameter $d{=}1.7\pm0.75$ $\mu$m, and axial length $H{=}6.7\pm0.86$ $\mu$m where the errors are standard deviations. These values agree with the reported numbers for \textit{B. sabtilis} and \textit{E. coli} \cite{Li2010,turner2000,darnton2007}. With the formula $H\sqrt{1+\pi^{2}(d/\lambda)^{2}}$ we obtain 12.13$\pm$2.3 $\mu$m as the contour length of the filaments~\cite{darnton2007}. 

Furthermore, we measured the three projected angles between the bundles in the front, middle, and back of the overlapping area by the bundles when they are oriented in the observation plane, as is schematically drawn in Fig.~\ref{fig:Fluorescence}g. These angles are correlated and show the same thing. They were measured together to ensure the tendency between different strains. Interestingly, the projected angles increase with the flagellar number of the cells which is shown in Fig.~\ref{fig:angles}. The middle projected angle between the bundles ranges from around 30$^{\circ}$ to 40$^{\circ}$, while a smaller angle would be expected in the three dimensional space.
We define an effective cell aspect ratio that includes the lateral extension of the bundles. This ratio depends on the position of the bundles, and for a constant opening angle, it is minimum (maximum) when the bundles depart from the body in the front (rear) of the cell (see Fig. S2). Cells with a few flagella can generally have a larger effective aspect ratio than the highly flagellated ones if the position of the bundles on the cell body differs approximately less than 1 $\mu$m (Fig. S2). The cell aspect ratio (including the bundles) of different strains ranges from $\simeq$1.5 to 4. An example of the different strains when the bundles are formed in the middle of the cell body is presented in Fig.~\ref{fig:angles}. The aspect ratio of the cell body is $\simeq$8, which is at least twice the effective aspect ratio, considering the flagellar bundles.\\          
Swimming speed, as a function of the opening angles for the wild type strain, is illustrated in Fig.~\ref{fig:Vr-a}. Corresponding correlation coefficients for the strains swrA$^{++}$, WT, and $\Delta$swrA are $r{=}-0.65$, $-0.42$, and $-0.15$, respectively. 

Swimming microorganisms are considered to be torque and force free. Therefore, the cell body rotates in the opposite direction of the flagellar bundle to compensate its torque~\cite{lauga2009}. The cell body is misaligned with the bundle and rotates around the velocity vector and bundle axis, which leads to a helical path of the body in space. Projected trajectories on the plane are sinusoidal, where the angle between the cell body and velocity direction is known as the wobbling angle. For the WT strain, the run speed increases with the decreasing wobbling angle (Fig.~\ref{fig:wobbling}a). Additionally, the wobbling angle increases for the strains with a higher flagellar number. A larger wobbling angle can originate from the wider opening angle between the bundles of the cells with a higher number of flagella. In general, the number of bundles is not very strongly correlated with the running speed, bundle width, and opening angle of the bundles for all the strains.  

% an example of a two-column table
%\begin{table*}
%\small
  %\caption{\ An example of a caption to accompany a table, table captions do not end in a full point}
  %\label{tbl:example}
  %\begin{tabular*}{\textwidth}{@{\extracolsep{\fill}}lllllll}
    %\hline
    %Header one & Header two & Header three & Header four & Header five & Header six  & Header seven\\
    %\hline
    %1 & 2 & 3 & 4 & 5 & 6  & 7\\
    %8 & 9 & 10 & 11 & 12 & 13 & 14 \\
    %15 & 16 & 17 & 18 & 19 & 20 & 21\\
    %\hline
  %\end{tabular*}
%\end{table*}

%%_____________________________________________________________________________________________________________________________________________________________________________________
\balance
\section{Discussion}

It has recently been reported that the flagellar number of \textit{B. subtilis} bacteria mainly affects the rotational diffusion and turning angle after tumbling~\cite{najafi2018}. A slight modification of these quantities can substantially influence the transport properties of the cells. The maximum measured turning angle for \textit{B. subtilis} strains is $\simeq$60$^{\circ}$, which is smaller than $\simeq$70$^{\circ}$ for \textit{E. coli} cells~\cite{berg1972,molaei2014,najafi2018}.
\textit{E. coli} is not polar, i.e., the bundle can be formed at either end of the cell body. This kind of bundle rearrangement leads to a large reorientation after tumbling. Due to the formation of multiple bundles in \textit{B. subtilis}, it is less possible that all of them reform on the opposite side of the body. Therefore, the average turning angle of \textit{B. subtilis} cells decreases, and they swim much more persistently than \textit{E. coli}.   
Consequently, the translational diffusion coefficient of \textit{E. coli} ranges from $\simeq$150 to 500 $\mu$m$^{2}$/s~\cite{vuppula2010,berg1990,lewus2001}, while the measured values for the strains in this study are between $\simeq$500 and 5000 $\mu$m$^{2}$/s~\cite{najafi2018}, which differ by one order of magnitude. Comparing the translational diffusion coefficients gives a robust insight into the cell motility since they can be calculated independent of the run-tumble characteristics of the swimmers.  

Common analytical models approximate the bundle as a rotating thick helix which is aligned with the main axis of the cell body. While this approach ignores the body wobbling, it predicts the swimming speed as a function of the bundle rotation rate and its geometrical properties~\cite{lauga2009,kanehl2014}.
Our experimental results suggest that the angle between the bundle and the body axis in the single bundle, or equivalently, the opening angle between multiple bundles, can be considered an additional geometrical feature, which can control several swimming parameters such as wobbling angle, rotational, and translational diffusion. The average wobbling frequency for all strains is about 9 Hz which agrees with 8 Hz reported for the swimming \textit{E. coli} cells close to the surface~\cite{bianchi2017}. Wobbling angle is correlated with the rotational diffusion (see Fig. S3) and both increase with the flagellar number~\cite{najafi2018}, which we associate to the widening of the opening angle between the bundles. The wobbling is due to the misalignment of the body relative to the bundle. The higher deviation of the bundle axis relative to the body direction and the likewise broader opening angle between the bundles eventually causes a larger wobbling angle. Increasing the angle between the bundles reduces the swimming speed in two different ways: (\romannumeral 1) it directly enhances the hydrodynamic resistance against swimming (approximately half of the translational drag of the whole cell can be due to its bundle~\cite{chattopadhyay2006}) (Fig.~\ref{fig:Vr-a}), and (\romannumeral 2) it enlarges the wobbling angle which rises the drag force on the cell body (Fig.~\ref{fig:wobbling}a).
The prediction of the swimming speed is a complex problem. The swimming speed depends on the position and alignment of the bundles relative to each other and the cell body, and the bundle rotation rate, which itself can be a function of bundle configuration. 
The arrangement of bundles spins CCW when it is observed from the rear of swimming direction due to the CW rotation of the cell body. The hydrodynamic force between the left-handed helices rotating in the CCW direction is attraction and they can collapse to a single bundle. It has been shown that only the rotation of the cell body is sufficient to wind the helices even in the absence of hydrodynamic interaction and helices rotation~\cite{powers2002}. However, the body length of \textit{B. subtilis} is longer than \textit{E. coli}, and the limited flagella length suppresses the collapse of bundles, as is shown in Fig.~\ref{fig:bundleextra}a, where short flagella cannot form a bundle.
The inclusion and entanglement of several flagella, from the broad area of the cell surface and into different bundles, can reduce the geometrical freedom of each bundle and keep them separated. The length of the overlapping area of the bundles anti-correlated with the opening angle between the bundles for each strain (see Fig. S4).
The overlapping area has the length of $\sim$4 $\mu$m, which is the same for all strains and is approximately half the length of body.
Therefore, it appears that the involvement of more flagella in the bundle is the main restriction which broadens the opening angle between the bundles.
Moreover, it qualitatively appears that longer cells have smaller wobbling angle as an evidence of persistent swimming due to the rather stabilized motility with the aid of lateral bundles. The contribution of the cell length on persistent swimming seems to be by the means of lateral bundles. This agrees with the recent report which shows that longer cephalexin treated \textit{E. coli} cells swim persistently~\cite{guadayol2017}.       
Similar functionality has also been observed for the secondary flagellar system of \textit{Shewanella}, in which the lateral flagella increases the directional persistency of cells~\cite{bubendorfer2014}. We conclude the appearance of multiple bundles in the swarmer cells are likely due to the cell elongation rather than hyper-flagellation~\cite{eisenstecken2016,kearns2010}.      

To summarize, we performed quantitative measurements of the swimming bacteria with multiple bundles and demonstrated that the number of bundles is independent of the flagellar number. The observation of three bundles is the most probable case due to the same body length of all the strains being the most basic geometrical factor.
We revealed that the projected angle between the bundles increases with the flagellar number, and accordingly, the effective aspect ratio of the cell decreases.
We qualitatively discussed how the larger angle between the bundles can indirectly reduce the swimming speed by widening the wobbling angle and increasing the drag force on the cell body.  
Our results highlight the impact of the swimmer geometry, especially the cell length and the relative alignment of bundles, as an additional freedom to adjust the motility parameters of the cells. The outcomes also provide further insights into the complicated dynamics of the highly flagellated and long bacterial species such as \textit{B. subtilis}.

\section*{Conflicts of interest} 
There are no conflicts of interest to declare.

\section*{Acknowledgment} 
We thank Daniel B Kearns for providing us with the strains. JN Acknowledges Emmanuel Terriac for his initial help with the fluorescence microscopy. JN is also grateful to Thomas Eisenstecken and Mahdiyeh Mousavi for a fruitful discussion. GB thanks the Deutsche Forschungsgemeinschaft (DFG) for support through the CRC-TRR 174.

%************check the correlation of wobbling angle(Dr) and cell length-->add correlation coefficients
%Dr and wobbling angle

%direct and indirect effect of bundle opening angles on speed
%bundle length is the same order of cell length
%************aspect ratio remarkably decreases with including the bundles
%citing and comparing with scientific report paper which was about persistency and cell length
%consequences of assymetric bundle arrangement
%*****why we can not compare the strains!!!???
%*****explain more the contribution of x and involvement of more flagella on opening angle

 %For footnotes in the main text of the article please number the footnotes to avoid duplicate symbols. e.g.  \footnote[num]{your text} the corresponding author \ast counts as footnote 1, ESI as footnote 2, e.g. if there is no ESI, please start at [num]=[2], if ESI is cited in the title please start at [num]=[3] etc. Please also cite the ESI within the main body of the text using \dag.


%\balance


\footnotesize{
\bibliography{rsc} %your .bib file

\bibliographystyle{rsc} %the RSC's .bst file
}

\end{document}